\documentclass[12pt,preprintnumbers,aps,amssymb,nofootinbib,amsmath]
{revtex4}
\usepackage{epsfig,epsf}
\usepackage{graphicx}
\usepackage{epstopdf}
\usepackage{bm} 
\newcommand{\beq}{\begin{equation}}
\newcommand{\eeq}{\end{equation}}
\newcommand{\bea}{\begin{eqnarray}}
\newcommand{\eea}{\end{eqnarray}}
\def\eq#1{{(\ref{#1})}}
\def\fig#1{{Fig.~\ref{#1}}}

\newcommand{\as}{\alpha_s}

\newcommand{\Tr}{{\rm Tr}}

\newcommand{\lagr}{\mathcal{L}}
\newcommand{\e}{\varepsilon}
\newcommand{\aver}[1]{\left\langle #1 \right\rangle}
\def\b#1{\mathbf{#1}}
\newcommand{\un}{\underline}
\begin{document}

\preprint{RBRC-658}


\title{
Gluon recombination in high parton density QCD: \\inclusive  pion production }

\author{Yang Li$\,^a$ and Kirill Tuchin$\,^{a,b}$\\}

\affiliation{
$^a\,$Department of Physics and Astronomy, Iowa State University, Ames, IA 50011 \\
$^b\,$RIKEN BNL Research Center,
Upton, NY 11973-5000\\}

\date{\today}


\begin{abstract}
We argue that the collinear factorization of the fragmentation functions in high energy hadron and nuclei collisions breaks down at transverse momenta $k_T\lesssim Q_s/g$  due to  high parton densities in the colliding hadrons and/or nuclei. We calculate, at next-to-leading order in projectile parton density and to all orders in target parton density,  the double-inclusive cross section for production of a pair of  gluons in the scalar $J^{PC}=0^{++}$ channel. Using the low energy theorems of QCD we find the inclusive cross section for  $\pi$-meson production.
\end{abstract}

\maketitle


\section{Introduction}\label{sec:intr}

Strong interactions at high energy are crucially influenced by high density of gluons in wave functions of colliding hadrons and/or nuclei \cite {GLR,Mueller:wy,Blaizot:nc}. A systematic description of particle production in high energy hadron reactions can be carried out in terms of quasi-classical solutions to the Yang-Mills equations on the light-cone \cite{MV} .
The quasi-classical approximation is valid for  field modes which occupation number reaches the saturation limit of $\sim 1/g$. These modes  correspond to gluons with transverse momentum less than the saturation momentum $Q_s$. The square of the saturation momentum is a measure of the two-dimensional color charge density in the hadron/nucleus wave function.
As the collision energy increases, the quantum fluctuations increase the occupation number of modes with larger transverse momentum which manifests  in increase of  $Q_s$. The corresponding high energy evolution of scattering amplitude is governed by the nonlinear Balitsky-Kovchegov evolution equation \cite{BK,JIMWLK}.  A comprehensive reviews of the physics of gluon saturation (Color Glass Condensate) can be found in Refs.~\cite{Iancu:2003xm,Jalilian-Marian:2005jf}.
A quasi-classical approach to inclusive processes of high energy QCD has driven a lot of attention due to its remarkable phenomenological success in small-$x$ DIS, $p(d)A$ and $AA$ collisions  \cite{Iancu:2003xm,Jalilian-Marian:2005jf,Bartels:2005yb,Tuchin:2006hz}.

One of the challenges of QCD is to understand how the color degrees of freedom metamorphose into hadrons. In the traditional perturbative QCD this problem is solved with the help of the collinear factorization theorems  which allow to separate the universal non-perturbative parton distribution functions, fragmentation functions, and the hard partonic sub-processes. The collinear factorization theorems hold only if particle production is characterized by a momentum scale which is much larger than the typical momentum scale in hadron wave functions. At high energies, existence of a semi-hard scale $Q_s$ -- which is an increasing function of energy and atomic mass $A$ -- does not allow application of factorization theorems at  transverse momenta of the order of $Q_s$ or smaller
\footnote{In fact, factorization theorems break down at even higher scale, leading to the so-called extended geometric scaling, see Ref.~\cite{Stasto:2000er,Levin:2000mv,Iancu:2002tr}.}. This is the kinematical region in which bulk of  particles is produced. Although the collinear factorization breaks down, the perturbation theory does not due to smallness of the coupling at the scale $Q_s$ even for the IR modes.
Failure of the collinear factorization of parton distribution functions has been discussed already in the pioneering publications  on gluon saturation \cite{GLR, Mueller:wy}. It was suggested that a more general type of factorization, $k_T$-factorization, may hold at high energies. However, it turned out that although the $k_T$-factorization is much better approximation of the exact formulas than the collinear factorization, it is violated as well in all high energy processes save for the single inclusive gluon production in $\gamma^*A$ collisions \cite{Kovchegov:2001sc}.

As the consequence of breakdown of factorization of the fragmentation functions,  hadronization pattern changes with parton density in both target and projectile and thus, exhibits complicated energy and atomic number dependence. In the present paper we discuss one of the possible hadronization channels which explicitly breaks down the collinear factorization.
From empirical point of view, failure of the collinear factorization is evident from observations of  strong energy  dependence of baryon to meson ratios in $pp$ and $dAu$ collisions \cite{Adams:2006nd}.

Another process in which the collinear factorization breaks down is  the  relativistic heavy-ion collisions in which  the produced system of color charges probably  evolves  through the dense and hot stage (Quark Gluon Plasma). However, it must be heeded that the collinear factorization is violated already at the early stages of decay of the classical fields, preceding the formation of the QGP. It has been suggested that parton recombination may be an alternative mechanism of hadronization \cite{Voloshin:2002wa,Fries:2003vb,Greco:2003mm,Hwa:2003ic,Molnar:2003ff}.  It yields a surprisingly good description of experimental data on elliptic flow  and might also be relevant for  interpretation of baryon to meson ratios reported in \cite{Adler:2003kg}.   It is based on a simple idea that   small-$k_T$ mesons are formed by  coalescence of two constituent quarks with transverse momenta  $k_T/2$ at the same rapidity, while baryons are formed by  coalescence of three constituent quarks with momenta $ k_T/3$ at the same rapidity.

Motivated by the phenomenological success of the recombination approach we
set to investigate the process of  gluon recombination in the high gluon density regime. In this paper we consider  pion production and argue that it is dominated by  the recombination of two classical fields. Our approach naturally incorporates  momentum conservation as well as the recombination geometry in the coordinate space (recombining partons must be in the same elementary volume of   phase-space). In heavy-ion collisions it yields a ``cold" nuclear matter effect on particle hadronization and may be essential for analysis of the residual ``hot" nuclear matter effect.

\begin{figure}[ht]
  \begin{center}
        \includegraphics[width=16cm]{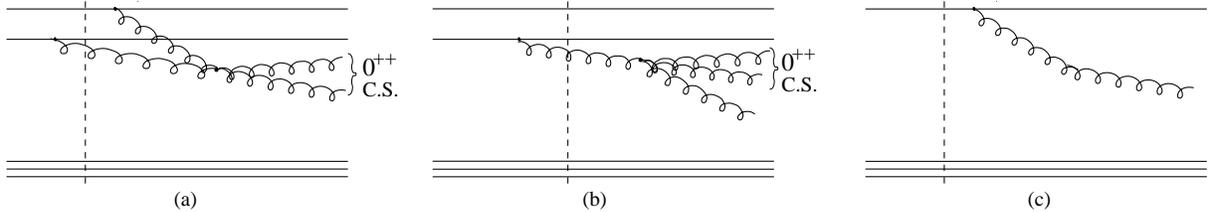}
\end{center}
\caption{Examples of diagrams contributing to the production of a pair of gluons in $J^{PC}=0^{++}$ and color singlet state: (a) quasi-classical case, (b) first radiative correction \cite{Kharzeev:1999vh,Kharzeev:2002rp,Kharzeev:2004ct}, (c) single gluon production followed by the conventional fragmentation \cite{Kovchegov:1998bi}. Horizontal solid lines are the valence quarks belonging to a different nucleons. Vertical dashed line describes an instantaneous interaction which can happen at different light cone times (for simplicity we show here only one  possible interaction; see \fig{wave_funct} for other possibilities).}
\label{compare}
\end{figure}

The recombination process which we discuss in the present paper consists of two stages: (i) production of a pair of gluons in $J^{PC}=0^{++}$ and color singlet state; (ii) recombination of this pair into a pair of $\pi$-mesons using the anomaly matching mechanism \cite{Kharzeev:1999vh}. The relevant diagrams are depicted in \fig{compare}.
Shown in \fig{compare}(a) is  emission of two gluons by  valence quarks belonging to  different nucleons in the incoming nucleus, their successive merging and production of a gluon pair in the singlet $J^{PC}=0^{++}$ state.

An alternative way to produce a pair of gluons in a singlet $J^{PC}=0^{++}$ state is shown in \fig{compare}(b). Unlike the diagram \fig{compare}(a) it represents a first step in quantum evolution which proceeds by emission of a soft gluon. This type of evolution gives rise to the Kharzeev--Levin soft Pomeron \cite{Kharzeev:1999vh} (see also \cite{Kharzeev:2002rp,Kharzeev:2004ct}). However, this  diagram is parametrically small as compared to \fig{compare}(a). Indeed, the diagram of \fig{compare}(a) is of the order of $\as^6\,\varrho_t\, \varrho_p^{2}\sim 1$, where $\varrho_t$ and $\varrho_p$ are the parton densities in the target and projectile respectively; $\varrho_t\sim A^{1/3}$, where
$A$ is the  atomic mass of the target nucleus  (if the projectile is a nucleus of atomic mass $B$, then $\varrho_p\sim B^{1/3}$). Here we used the fact that in a quasi-classical (McLerran-Venugopalan) approximation $\as^2\,\varrho_t\sim \as^2\,\varrho_p\sim 1$. On the other hand, the diagram \fig{compare}(b) is parametrically of the order of $\as^5\, \varrho_t\,\varrho_p\sim \as$.

At low invariant masses, description of pair production in the singlet $J^{PC}=0^{++}$ channel in terms of color degrees of freedom becomes inadequate. Spectral density of the corresponding correlator (see Sec.~\ref{sec:3}) is saturated by  colorless excitations the most prominent of which are  pions. Unlike gluons which contribute to the spectral density at the order $\as^2$, see \eq{rho.pert}, pions contribute  at the order $\as^0$, see\eq{rhopipi}. Therefore,
 diagram \fig{compare}(a)  is of the order of $\as^4\,\varrho_t\,\varrho_p^2\sim 1/\as^2$ at low invariant masses. This is parametrically larger than the hadron production via the collinear fragmentation of a single gluon  \cite{Kovchegov:1998bi,Braun:2000bh,Kovchegov:2001sc,Dumitru:2001ux,Blaizot:2004wu} shown in \fig{compare}(c). Indeed, the corresponding single gluon production diagram is of the order of $\as^3 \,\varrho_t\,\varrho_p\sim 1/\as$. On the other hand, diagram \fig{compare}(a) represents a higher twist effect as compared to the single gluon production and thus has an additional suppression factor $Q_s^2/k_T^2$ at high transverse momenta (here $Q_s$ is associated with the projectile). These two hadronization  processes become of the same order at transverse momenta of the order of $ k_T^2\sim Q_s^2/\as$. We expect that at lower transverse momenta the recombination mechanism discussed in this paper gives the main contribution to the particle hadronization at high energies. At RHIC energies, the corresponding kinematic region is about $k_\bot\lesssim 3$~GeV for light hadrons in Au-Au collisions at midrapidity.  It is significantly wider at forward rapidities and at higher energies.

The paper is organized as follows. In Sec.~\ref{sec:2} we calculate the double inclusive cross section for gluon pair  production in the singlet $J^{PC}=0^{++}$ channel in the framework of the dipole model  \cite{dipole}. Since the relevant degrees of freedom at low invariant masses are pions which have large inelastic cross section on a nucleon,  we are going to neglect the diagram \fig{wave_funct}D in which the produced pair interacts with the target. Indeed, the survival probability of a pion in a  heavy nucleus  is exponentially suppressed as compared to the one of a color dipole which has much smaller characteristic size $\sim 1/Q_s$. This approximation amounts to assumption that the intermediate gluons in \fig{wave_funct} are almost on-mass-shell, see \eq{gauge2}.  The resulting ``wave function" is given by \eq{wf-z-half}. We then calculate the forward amplitude of each color dipole in a nucleus,   \eq{scattt} and that  of the entire projectile \eq{mres}. In doing that we neglect correlation between partons belonging to  different nucleons since the corresponding dipoles have sizes about $1.3$~fm (typical nucleon separation).
The double inclusive cross section of gluon pair production is the convolution of the projectile ``wave function" and the forward scattering amplitude  in the coordinate space and is given by \eq{XSECT3}.

In Sec.~\ref{sec:3}, following in steps of \cite{Kharzeev:1999vh}, we describe the formalism of anomaly matching and calculate the double inclusive cross section for $\pi$-meson production, \eq{main}.
Our approach is based on the observation that the scale anomaly is
closely related to the finite density of vacuum fluctuations of quantum gluon fields. These fluctuations are characterized by a semi-hard scale $M_0\simeq 2-2.5$~GeV \cite{Novikov:1981xj,Fujii:1999xn}. The presence of this scale makes reasonable the perturbative expansion and allows for the calculation of  non-perturbative contributions to the spectral density due to the scale anomaly.  Our method is similar to the QCD sum rules approach. The later  is  based on the operator product expansion which holds only in the presence of a hard scale.

In Sec.~\ref{sec:evol} the quantum evolution effects are taken into account. This makes it possible to address  energy, rapidity and atomic mass dependence of the  pion production cross section. The final result is given in Eq.~\eq{main_ev}.

The fragmentation process suggested in this paper has a number of phenomenological consequences. Firstly, the energy dependence of the cross section is steeper than in the conventional fragmentation mechanism since it requires exchange of an additional Pomeron between the projectile and the produced pair. Secondly, as \fig{Mspec:fig} implies, at higher  energies  heavier hadrons can be produced along with pions. Therefore this hadronization mechanism yields a non-trivial energy dependence for different particle species and can be used to analyze the energy dependence of particle ratios.
This issue is addressed in a great detail in the forthcoming publication \cite{FP}. Other two-gluon channels besides the $J^{PC}=0^{++}$ may also have interesting phenomenological applications and can be considered along the same lines.



\section{Two-gluon production in a quasi-classical approximation}\label{sec:2}
\begin{figure}
    \begin{center}
        \includegraphics[width=16cm]{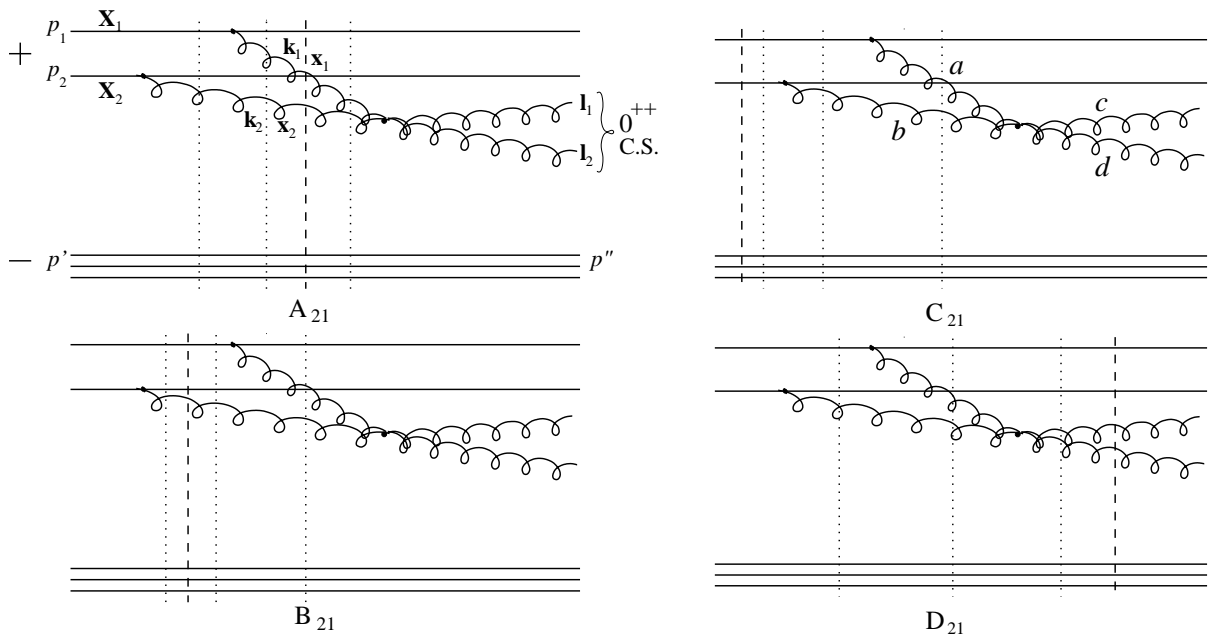}
\end{center}
\caption{Contributions to production of a gluon pair in $J^{PC}=0^{++}$ and color singlet state in a quasi-classical approximation. Only diagrams in which the gluon at $\b x_2$ is emitted from a valence quark before the gluon at $\b x_1$ are shown.}
\label{wave_funct}
\end{figure}

In the light-cone perturbation theory \cite{Lepage:1980fj} there are eight diagrams contributing to the ``wave function" of a gluon pair. In \fig{wave_funct} four of them are shown; the other four diagrams can be obtained by switching the order of gluon emission from \emph{valence quarks}. Note that in the eikonal approximation, the life-time of a parton fluctuation in the fast nucleus wave function $t_\mathrm{f}\simeq k_+/\b k^2$ is much larger than the typical time of interaction with the target $t_\mathrm{i}\simeq R_A$, where $k_+$ is the large light-cone momentum of the parton, $\b k$ its transverse momentum and $R_A$ is the nuclear radius. Therefore, in this approximation we can regard the interaction as instantaneous. This observation constitutes the basis of the dipole model \cite{dipole}. It implies that an inclusive production cross section is a convolution  of a projectile ``wave function" with the dipole scattering amplitude in the transverse coordinate space.

Diagrams in \fig{wave_funct} differ one from another by the structure of their energy denominators which we would like  to consider now more closely.
Each of the eight diagrams   in \fig{wave_funct} contains three energy denominators.
In the diagrams A--C the two leftmost energy denominators corresponding to emission of gluons at $\b x_1$ and $\b x_2$ are different for each diagram whereas  the rightmost one, corresponding to the 2$\to$2 gluon scattering process  is the same in all cases. The product of the first two energy denominators is given by
\begin{subequations}\label{denoms}
\bea\label{denomA}
&&A_{21}+A_{12}:\quad \frac{1}{k_{2-}}\,\frac{1}{k_{1-}+ k_{2-}} + \frac{1}{k_{1-}}\,\frac{1}{k_{1-}+k_{2-}}=\frac{1}{k_{1-}}\,\frac{1}{k_{2-}}\,,\\
&&\label{denomB1}
B_{21}:\quad \frac{1}{k_{2-}}\,\frac{1}{k_{2-}+p_-''-p_-'}=\frac{1}{k_{2-}}\,\frac{1}{k_{2-}-l_{1-}-l_{2-}}\,,\\
&&\label{denomB2} B_{12}: \quad \frac{1}{k_{1-}}\,\frac{1}{k_{1-}+p_-''-p_-'}=\frac{1}{k_{1-}}\,\frac{1}{k_{1-}-l_{1-}-l_{2-}}\,,\\
\label{denomC1}
&&C_{21}:\quad \frac{1}{p''_--p'_-}\,\frac{1}{k_{2-}+p''_--p'_-}=-\frac{1}{l_{1-}+l_{2-}}\,\frac{1}{k_{2-}-l_{1-}-l_{2-}}\,,\\
&&\label{denomC2}C_{12}:\quad \frac{1}{p''_--p'_-}\,\frac{1}{k_{1-}+p''_--p'_-}=-\frac{1}{l_{1-}+l_{2-}}\,\frac{1}{k_{1-}-l_{1-}-l_{2-}}\,,
\eea
while the last energy denominator  is
\beq
\frac{1}{k_{1-}+k_{2-}+p_-''-p_-'}=\frac{1}{k_{1-}+k_{2-}-l_{1-}-l_{2-}}\,,
\eeq
where we used the overall energy conservation condition $p_-''+l_{1-}+l_{2-}=p_-'$. The product of the energy denominators in the diagram D is
\beq
\label{denomD}
D_{21}+D_{12}: \quad \left(\frac{1}{k_{1-}}+\frac{1}{k_{2-}}\right)\,\frac{1}{k_{1-}+k_{2-}}\,\frac{1}{l_{1-}+l_{2-}}=\frac{1}{k_{1-}k_{2-}}\,\frac{1}{l_{1-}+l_{2-}}\,.
\eeq
\end{subequations}
Simple calculation shows that sum of all energy denominators from all eight diagrams vanishes. We conclude  that in the  absence of interactions
\beq\label{gauge}
\sum_{i,j=1,2\,, i\neq j}\left(\Psi_{A_{ij}}+\Psi_{B_{ij}}+\Psi_{C_{ij}}+\Psi_{D_{ij}}\right)=0
\eeq
as expected.

As explained in the Introduction, the correct description of the particle pair produced at low invariant masses in the scalar $J^{PC}=0^{++}$  state is furnished using the hadronic degrees of freedom. By virtue of  the color transparency, pion-nucleon inelastic cross section is much larger than the dipole-nucleon one. Hence, we are tempted to neglect the contribution of diagram D to inclusive cross section.
This  imposes a certain constraint on the energy denominators A--C since we must ensure that the condition \eq{gauge} holds, for otherwise the resulting cross section would not be gauge invariant. In other words, we require that
\beq\label{gauge2}
\sum_{i,j=1,2\,, i\neq j}\left(\Psi_{A_{ij}}+\Psi_{B_{ij}}+\Psi_{C_{ij}}\right)\approx 0\,.
\eeq
This approximate equation must hold with the same accuracy as the assumption that pions are completely absorbed by the nucleus. It can be guaranteed if the intermediate gluons are almost on mass-shell, i.\ e.\ $k_{1-}+k_{2-}\approx l_{1-}+l_{2-}$. Then, Eqs.~\eq{denoms} become
\begin{subequations} \label{denoms2}
\beq\label{denomAa}
A_{21}+A_{12}:\quad \frac{1}{k_{2-}}\,\frac{1}{k_{1-}+k_{2-}}+\frac{1}{k_{1-}}\,\frac{1}{k_{1-}+k_{2-}}=\frac{1}{k_{1-}}\,\frac{1}{k_{2-}}\,,
\eeq
\beq\label{denomBb}
B_{21}:\quad \frac{1}{k_{2-}}\,\frac{1}{p_-''+k_{2-}-p_-'}\approx -\frac{1}{k_{1-}}\,\frac{1}{k_{2-}}\,,\quad
B_{12}:\quad
\frac{1}{k_{1-}}\,\frac{1}{p_-''+k_{1-}-p_-'}\approx -\,\frac{1}{k_{1-}}\,\frac{1}{k_{2-}}\,,
\eeq
\beq\label{denomCc}
C_{21}+C_{12}:\quad \frac{1}{p''_--p'_-}\,\frac{1}{k_{2-}+p''_--p'_-}+\frac{1}{p''_--p'_-}\,\frac{1}{k_{1-}+p''_--p'_-}\approx \frac{1}{k_{1-}}\,\frac{1}{k_{2-}}\,.
\eeq
\end{subequations}
Obviously, sum of all energy denominators in \eq{denoms2} conforms to condition \eq{gauge2}.

Since the structure of the energy denominators in cases A--C is the same, we need to calculate the ``wave function" in only one case, say,  A. Introducing  momentum  $q$ such that (see \fig{wave_funct})
\beq\label{k1k2}
k_1=\frac{1}{2}( l_1+ l_2)-q\,,\quad k_2=\frac{1}{2}( l_1+ l_2)+ q\,,
\eeq
we have
\bea\label{wfA}
\Psi^{\lambda_3\lambda_4}(l_1,l_2) &=&T_a\,T_b\, g^2\sum_{\lambda_1\lambda_2}\,
\int \frac{\bar u(p_1)}{\sqrt{p_{1+}}}\gamma\cdot\e^{\lambda_1*}\frac{u(p_1-(l_1+l_2)/2+q)}{\sqrt{(p_1-(l_1+l_2)/2+q)_+}}\,
 \frac{1}{k_{1-}}\nonumber\\
 &&\times
\frac{\bar u(p_2)}{\sqrt{p_{2+}}}\gamma\cdot\e^{\lambda_2*}\frac{u(p_2-(l_1+l_2)/2-q)}{\sqrt{(p_2-(l_1+l_2)/2-q)_+}}\frac{1}{k_{2-}}\nonumber\\
&&\times
\frac{1}{k_{1-}+k_{2-}-l_{1-}-l_{2-}-i0}\,\e^{\lambda_1}_\mu\e^{\lambda_2}_\nu\Gamma^{\mu\nu\rho\sigma}_{abcd}\aver{\e^{\lambda_3*}_\rho\e^{\lambda_4*}_\sigma }_S\,\delta_{cd}\nonumber\\
&&\times
\frac{1}{((l_1+l_2)/2-q)_+}\,
 \frac{1}{((l_1+l_2)/2+q)_+}\,\frac{d^3q}{16\pi^3}\,,
\eea
where $\Gamma^{\mu\nu\rho\sigma}_{abcd}$ is the four-gluon vertex:
\beq\label{4gluon}
\Gamma^{\mu\nu\rho\sigma}_{abcd}=g^2[f^{abe}f^{cde}(g^{\mu\rho}g^{\nu\sigma}-g^{\mu\sigma}g^{\nu\rho})+f^{ace}f^{bde}(g^{\mu\nu}g^{\rho\sigma}-g^{\mu\sigma}g^{\nu\rho})+
f^{ade}f^{bce}(g^{\mu\nu}g^{\rho\sigma}-g^{\mu\rho}g^{\nu\sigma})]\,,
\eeq
and
$\aver{\ldots}_S$ means projection onto the singlet state. After projection of the final gluons onto the color singlet state, the color factor becomes the same for all six terms contributing to the four-gluon vertex.

In the eikonal approximation, we can simplify the $q\to qg$ vertex as follows \cite{dipole}
\beq\label{curr}
\frac{\bar u(p_1)}{\sqrt{p_{1+}}}\gamma\cdot\e^{\lambda_1*}\frac{u(p_1-k_1)}{\sqrt{(p_1-k_1)_+}}\,
 \frac{1}{k_{1-}}\approx \frac{2\,\b k_1\cdot \b e^{\lambda_1*}}
 {\b k_1^2}\,,
 \eeq
 where the bold typeface distinguishes the transverse component of the corresponding four-vector.
It is convenient to introduce the following notations: (i) fractions $z$ and $z'$ of the light-cone momenta of each of the four gluons are given by
\begin{subequations}
\beq
k_{1+}\equiv\frac{l_{1+}+l_{2+}}{2}-q_+=z\,(l_{1+}+l_{2+})\,,\quad
k_{2+}\equiv\frac{l_{1+}+l_{2+}}{2}+q_+=(1-z)\,(l_{1+}+l_{2+})\,.
\eeq
\beq
l_{1+}=z'\,(l_{1+}+l_{2+})\,,\quad l_{2+}=(1-z')\, (l_{1+}+l_{2+})\,;
\eeq
\end{subequations}
(ii) the total and the relative transverse momenta $\b k$, $\bm \kappa$ and $\tilde {\b q}$ are defined as
\beq\label{kappa}
\b k=\b l_1+\b l_2=\b k_1+\b k_2\,,\quad
\bm\kappa=z'\,\b l_2-(1-z')\,\b l_1\,,\quad
\tilde {\b q}=z\,\b k_2-(1-z)\,\b k_1\,.
\eeq
Note that the invariant mass of the produced pair is
\beq\label{inv.mass}
M^2=(l_1+l_2)^2=l_{1+}l_{2-}+l_{1-}l_{2+}-2\,\b l_1\cdot \b l_2=
\frac{1}{z'(1-z')}\,\bm \kappa^2\,.
\eeq
With the help of these equations, after some simple algebra, we can derive
\bea\label{2to2a}
&&\frac{1}{k_{1-}+k_{2-}-l_{1-}-l_{2-}-i0}=\frac{1}{\frac{\b k_1^2}{zk_+}+\frac{\b k_2^2}{(1-z)k_+}-\frac{\b l_1^2}{z'k_+}-\frac{\b l_2^2}{(1-z')k_+}-i0}
\nonumber\\
&=&\frac{k_+\,z\,(1-z)\,z'\,(1-z')}
{\tilde{\b q}^2 \, z' (1-z') - \bm\kappa^2 z(1-z)-i0}\,.
\eea
(We added expression $-(\b k_1+\b k_2)^2/k_++(\b l_1+\b l_2)^2/k_+= 0$ to the denominator of the first line).

At high energy, the scattering matrix is diagonal with respect to the color dipoles. Therefore, in order to take the interaction of the incoming parton system with the nucleus into account we need to transform the wave function $\Psi(l_1,l_2)$ into the coordinate representation. This is accomplished as follows:
 \beq\label{acc}
 \Psi(\bm \xi_1,\bm \xi_2)=\int\frac{d^2l_1}{(2\pi)^2}\int\frac{d^2l_1}{(2\pi)^2}
 e^{-i\,\b l_1\cdot \bm \xi_1-i\,\b l_2\cdot \bm \xi_2}\, \Psi(\b l_1, \b l_2)\,.
\eeq
Here $\bm \xi_1$ and $\bm \xi_2$ are the coordinates of the final gluons with momenta $\b l_1$ and $\b l_2$, respectively. It is convenient to introduce the transverse  coordinates of the intermediate gluons $\b x_1$ and $\b x_2$.
To this end, we change the set of integration variables in \eq{acc} and \eq{wfA} from $\{\b l_1, \b l_2, \b q\}$ to a new one $\{\b k, \bm \kappa,\b q\}$
using \eq{k1k2} and \eq{kappa}. The Jacobian of this transformation is unity.
The phase factor in \eq{acc} becomes
\beq
-i\,\b l_1\cdot \bm \xi_1-i\,\b l_2\cdot \bm \xi_2=-i\,\bm \kappa\cdot (\bm \xi_2-\bm \xi_1)-i\,
(\b k_1+\b k_2)\cdot (z'\,\bm \xi_1+(1-z')\,\bm \xi_2)\,.
\eeq
We can identify $ \b x_1=\b x_2 =z'\,\bm \xi_1+(1-z')\, \bm \xi_2\equiv\b x$. We observe that the coordinates of both intermediate gluons are equal. This is the result of the integration over the internal momentum $\b q$. Indeed, had we Fourier transformed the amplitude with respect to $\b l_1$, $\b l_2$ \emph{and} $\b q$,  the subsequent integration over $\b q$ would have given the delta function $\delta(\b x_2-\b x_1)$.

Summation over the gluon polarizations can be carried out using the rule
$\aver{\b e^{\lambda}_i\b e^{\lambda *}_j}=\delta_{ij}/2$. Recalling that in the light-cone gauge $\e\cdot \eta=\e_+=0$ we derive using \eq{4gluon}
\bea\label{pol.vec}
&&\sum_{\lambda_1,\lambda_2}\,\e^{\lambda_1}_\mu\e^{\lambda_2}_\nu\Gamma^{\mu\nu\rho\sigma}_{abcd}\aver{\e^{\lambda_3*}_\rho\e^{\lambda_4*}_\sigma}_S \,
(\b k_1\cdot \b e^{\lambda_1*})\,(\b k_2\cdot \b e^{\lambda_2*})\,\delta_{cd}
\nonumber\\
&=& -\,g^2\sum_{\lambda_1,\lambda_2} \,(\b e^{\lambda_1}\cdot \b e^{\lambda_2})\,
[f^{ace}f^{bde}+f^{ade}f^{bce}]\,\delta_{cd}
 \,(\b k_1\cdot \b e^{\lambda_1*})\,(\b k_2\cdot \b e^{\lambda_2*})\,
\delta_{\lambda_3\lambda_4} \nonumber\\
 &=& -\,g^2\,(\b k_1\cdot \b k_2)\,[f^{ace}f^{bde}+f^{ade}f^{bce}]\,\delta_{cd}\,\delta_{\lambda_3\lambda_4}
 =-g^2\,(\b k_1\cdot \b k_2)\,2\,N_c\,\delta_{ab}\,\delta_{\lambda_3\lambda_4}\,.
\eea

We can now write the ``wave function" as
\bea\label{wfA2}
\Psi^{\lambda_3\lambda_4}(\bm \xi,\b x)&=& -\, g^4\,T_a\,T_b\,2N_c\,\delta_{ab}\, \delta_{\lambda_3\lambda_4}\,\int\frac{d^2k}{(2\pi)^2}\int\frac{d^2\kappa}{(2\pi)^2}e^{-i\,\bm \kappa\cdot \bm \xi-i\,\b k\cdot \b x}
\int \frac{d^2\tilde q}{16\pi^3}\int_0^1 dz\nonumber\\
&\times&4\,\frac{(z\,\b k-\tilde{\b q})\cdot ((1-z)\,\b k+\tilde{\b q})}{(z\,\b k-\tilde{\b q})^2\,((1-z)\,\b k+\tilde{\b q})^2}\,
 \frac{z'(1-z')}{\tilde{\b q}^2 \, z' (1-z') - \bm\kappa^2 z(1-z)-i0}\,,
\eea
where $\bm \xi=\bm \xi_2-\bm \xi_1$.

The cross section for the double-inclusive gluon production is given by
\bea\label{xsection}
&&\frac{d\sigma_{gg}}{d^2l_1d^2l_2dyd^2b}=\frac{1}{8N_c^2}\frac{1}{(16\pi^3)^2}\,\Tr_1\Tr_2\,\sum_{\lambda_3\lambda_4}\int d^2\xi\int d^2x\, \int d^2\eta\,\int d^2y\,\int\frac{dz'}{z'(1-z')}\nonumber\\
&&\times\Psi^{\lambda_3\lambda_4}(\bm \xi,\b x)\Psi^{\lambda_3\lambda_4*}(\bm \eta,\b y)\,2\,\Xi(\b x,\b y)\, e^{i\,\b l_1\cdot\left(\b x-\b y+(1-z')\,(\bm \xi-\bm \eta)\right)}\,e^{i\,\b l_2\cdot\left(\b x-\b y-z'\, (\bm \xi-\bm \eta)\right)}\,,
\eea
where the coordinates $\b y$ and $\bm \eta$ in the complex-conjugate amplitude correspond to the coordinates $\b x$ and $\bm\xi$ in the amplitude; 
$\Xi(\b x,\b y)$ is the rescattering factor, $\b b$ is impact parameter of the projectile with respect to the target and  the color traces are taken over each nucleon (hence the subscripts 1,2).
Introducing color gluon notations in the amplitude as shown in  \fig{wave_funct} and denoting by letters with bars the corresponding gluon colors in the complex-conjugated amplitude, we   find  the overall color factor as
\bea\label{col_f}
&&\frac{1}{N_c^2}\Tr(T_aT_{\bar a})\,\Tr (T_bT_{\bar b})\,(f^{ace}f^{bde}+
f^{ade}f^{bce})\,\delta^{cd}\,
(f^{\bar a\bar c\bar e}f^{\bar b\bar d\bar e}+
f^{\bar a\bar d\bar e}f^{\bar b\bar c\bar e})\,\delta^{\bar c\bar d}\nonumber\\
&=& \frac{1}{4N_c^2}\,\delta^{a\bar a}\, \delta^{b\bar b}\,(N_c\, \delta^{ab}+N_c\delta^{ab})\,\,(N_c\, \delta^{\bar a\bar b}+N_c\delta^{\bar a\bar b})= 2\,C_FN_c\,.
\eea
We can see that all four non-vanishing gluon indices permutations in the four-gluon vertex give the same color factor.

Next, we turn to calculation of the forward scattering amplitudes for each time and gluon ordering. Let us for a moment assume that the intermediate gluons have an arbitrary coordinates $\b x_1$ and $\b x_2$ in the amplitude and $\b y_1$ and $\b y_2$ in the complex conjugate one.
\begin{figure}[ht]
    \begin{center}
      \includegraphics[width=15cm]{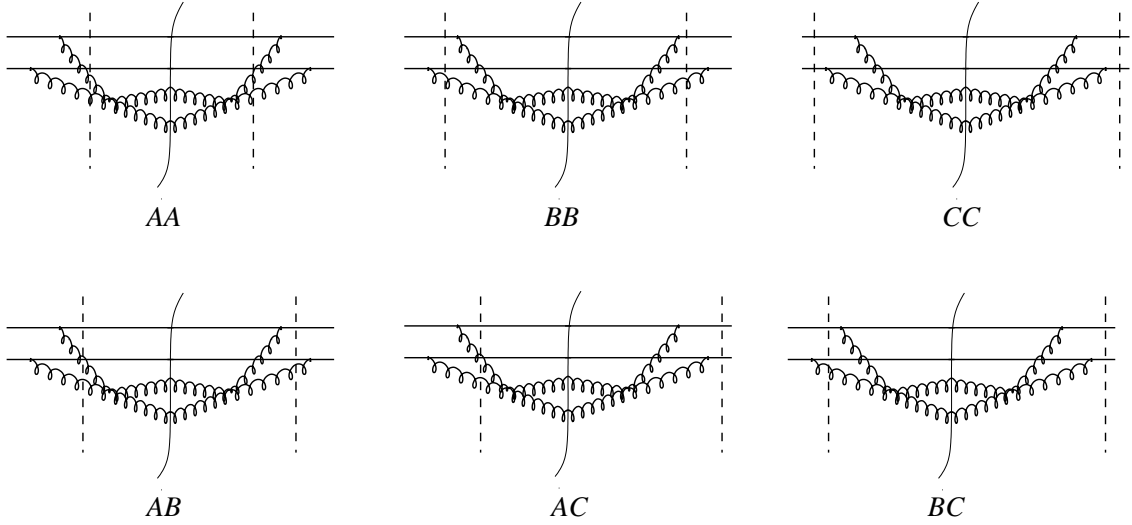}
\end{center}
\caption{All possible time sequences for interaction of a projectile with the target. The diagrams BA, CA and CB can be obtained by complex-conjugation of the diagrams AB, AC and BC respectively. Interaction of the final gluon pair is neglected as explained in the text.}
\label{scatter_fig}
\end{figure}
Calculation of the scattering amplitudes is similar to the case of a single gluon production\cite{Kovchegov:1998bi}. We have
\begin{subequations}\label{scattt}
\bea
\Xi_{AA}&=&e^{-\frac{1}{4}(\b x_1-\b y_1)^2 Q_s^2}\,e^{-\frac{1}{4}(\b x_2-\b y_2)^2 Q_s^2}\,,\\
\Xi_{BB}&=&e^{-\frac{1}{4}(\b x_1-\b y_1)^2 Q_s^2}+e^{-\frac{1}{4}(\b x_2-\b y_2)^2 Q_s^2} +e^{-\frac{1}{4}(\b x_1^2+\b y_2^2) Q_s^2}+e^{-\frac{1}{4}(\b x_2^2+\b y_1^2) Q_s^2}\,,\\
\Xi_{CC}&=&1\,,\\
\Xi_{AB}&=& e^{-\frac{1}{4}\b x_1^2 Q_s^2}\, e^{-\frac{1}{4}(\b x_2-\b y_2)^2 +Q_s^2}+e^{-\frac{1}{4}\b x_2^2 Q_s^2}\, e^{-\frac{1}{4}(\b x_1-\b y_1)^2 +Q_s^2}\,,\\
\Xi_{AC}&=& e^{-\frac{1}{4}\b x_1^2 Q_s^2}\,e^{-\frac{1}{4}\b x_2^2 Q_s^2}\,,\\
\Xi_{BC}&=& e^{-\frac{1}{4}\b x_1^2 Q_s^2}+e^{-\frac{1}{4}\b x_2^2 Q_s^2}\,.
\eea
\end{subequations}
Taking into account  Eqs.~\eq{denomAa}-\eq{denomCc} we can sum up all the scattering amplitudes to obtain
$$
\Xi_{AA}+\Xi_{BB}+\Xi_{CC}-\Xi_{AB}-\Xi_{BA}+\Xi_{AC}+\Xi_{CA}-\Xi_{BC}+\Xi_{CB}=
$$
\beq\label{rescatt}
\left(1+e^{-\frac{1}{4}(\b x_1-\b y_1)^2 Q_s^2}- e^{-\frac{1}{4}\b x_1^2 Q_s^2}-e^{-\frac{1}{4}\b y_1^2 Q_s^2}\right)
\left(1+e^{-\frac{1}{4}(\b x_2-\b y_2)^2 Q_s^2}- e^{-\frac{1}{4}\b x_2^2 Q_s^2}-e^{-\frac{1}{4}\b y_2^2 Q_s^2}\right)\,.
\eeq
In the limit of recombining gluons $\b x_1=\b x_2=\b x$ and $\b y_1=\b y_2=\b y$ we derive
\beq\label{mres}
\Xi(\b x,\b y)=\left(1+e^{-\frac{1}{4}(\b x-\b y)^2 Q_s^2}- e^{-\frac{1}{4}\b x^2 Q_s^2}-e^{-\frac{1}{4}\b y^2 Q_s^2}\right)^2\,.
\eeq

To write the final expression for the double-inclusive gluon production, we introduce an auxiliary function $F( \bm\xi ,\b x)$ such that
\beq\label{F}
\Psi^{\lambda_3\lambda_4}(\bm \xi,\b x)=-\, g^4\,T_a\,T_b\,2N_c\,\delta_{ab}\, \delta_{\lambda_3\lambda_4}\,F(\bm \xi,\b x)\,.
\eeq
With this notation we obtain
\bea\label{xsect}
&&\frac{d\sigma_{gg}}{d^2l_1\,d^2l_2\,dy\,d^2b}=\frac{\as^4}{2\pi^2}\,N_cC_F\int d^2\xi \int d^2x\, \int d^2\eta\,\int d^2y\,\int\frac{dz'}{z'(1-z')}\nonumber\\
&&\times F(\bm \xi ,\b x)\,F^*(\bm \eta,\b y)\,2\,\Xi(\b x,\b y)\, e^{i\,\b l_1\cdot\left(\b x-\b y+(1-z')\,(\bm \xi-\bm \eta)\right)}\,e^{i\,\b l_2\cdot\left(\b x-\b y-z'\,(\bm \xi-\bm \eta)\right)}\,.
\eea

Eq.~\eq{xsect} is a general result which we derived in the eikonal approximation and assuming that the recombining gluons are almost on-mass-shell. It can be significantly simplified if we  note that the recombining gluons must be close in rapidity which implies that the light-cone momentum fractions carried by gluons are typically equal $z,z'\approx 1/2$. Assuming that this configuration is dominant, we derive in Appendix~A the following expression for the ``wave function"
\beq\label{wf-z-half}
\Psi^{\lambda_3\lambda_4}(\bm \xi,\b x)=T_a\,T_b\,2N_c\,\delta_{ab}\, \delta_{\lambda_3\lambda_4}\,\frac{\as^2}{\pi^2\, x^2}\,\ln\left( 1-\frac{4x^2}{\xi^2}\right)
\eeq
and the cross-section
\bea\label{xsect2}
\frac{d\sigma_{gg}}{d^2l_1\,d^2l_2\,dy\,d^2b}&=&\frac{2\as^4}{\pi^6}\,\frac{N_cC_F}{(\b l_2-\b l_1)^4}\,
\int \frac{d^2x}{x^2}\int\frac{d^2y}{y^2}\,
\big[1-ix|\b l_2-\b l_1|\, K_1(i x |\b l_2-\b l_1|)\big]\,\nonumber\\
&\times&
\big[1+iy|\b l_2-\b l_1|\, K_1(-i y |\b l_2-\b l_1|)\big]
e^{i(\b l_1+\b l_2)\cdot (\b x-\b y)}\, 2\,\Xi(\b x, \b y)\,.
\eea

Using \eq{xsect2}  we can extract the inclusive cross section for production of a pair of  gluons with invariant mass $M$. For this objective, it is convenient to consider the inclusive cross section in terms of momenta $\b k=\b l_2+\b l_1$ and $\bm \kappa=(\b l_2-\b l_1)/2$. Then integrating over the directions of $\bm\kappa$ and  recalling that by \eq{inv.mass} $\bm\kappa^2=M^2/4$ yields
\bea\label{XSECT3}
\frac{d\sigma_{gg}}{dM^2\,d^2k\,dy\,d^2b}&=&\frac{8\as^4}{\pi^7}\,\frac{N_cC_F}{M^4}\,
\int \frac{d^2x}{x^2}\int\frac{d^2y}{y^2}\, e^{i\b k\cdot (\b x-\b y)}
\,\nonumber\\
&\times&
\big[1-ixM\, K_1(i x M)\big]\,\big[1+iyM\, K_1(-i y M)\big]
\, 2\,\Xi(\b x, \b y)\,.
\eea
Integrating\footnote{Convergence of the integral in \eq{XSECT3} is discussed in Appendix~C.} over $\b k$  we get
\beq\label{M-spec}
\frac{d\sigma_{gg}}{dM^2\,d^2b}=\frac{32}{\pi^6}\frac{N_cC_F\as^4}{M^4}\,\int_0^\infty\frac{dx}{x^3}\, \big|1-ixM\, K_1(i x M)\big|^2\,8\,\left(1-e^{-\frac{1}{4}x^2Q_s^2}\right)^2\,.
\eeq
The integrand of \eq{M-spec} can be expressed in terms of Bessel functions exploiting the relation
\beq
\big|1-ixM\, K_1(i x M)\big|^2=\left( 1+\frac{\pi}{2}\, M x\,Y_1(Mx)\right)^2+\frac{\pi^2}{4}\,(Mx)^2\, J_1^2(Mx)\,.
\eeq

In the limit of  large invariant masses $M\gg Q_s$ the dominant contribution to the integral in \eq{M-spec} comes from dipoles of size $1/M\ll x\ll 1/Q_s$. Expanding both the McDonald function and the dipole scattering amplitude yields
\beq\label{mdist1}
\frac{d\sigma_{gg}}{dM^2\,d^2b}\approx \frac{8\as^4 N_cC_F}{3\pi^5}\,\frac{Q_s}{M^3}\,,\quad M\gg Q_s\,.
\eeq
In the opposite limit of small invariant masses the largest (logarithmic) contribution stems from sizes $1/Q_s\ll x\ll 1/M$ in which case we estimate
\beq\label{mdist2}
\frac{d\sigma_{gg}}{dM^2\,d^2b}\approx \frac{64\as^4 N_cC_F}{\pi^6}\,\frac{1}{M^2}\,,\quad M\ll Q_s\,.
\eeq
This behavior has important phenomenological consequences which will be elucidated in the upcoming publication \cite{FP}.

\begin{figure}[ht]
    \begin{center}
      \includegraphics[width=8cm]{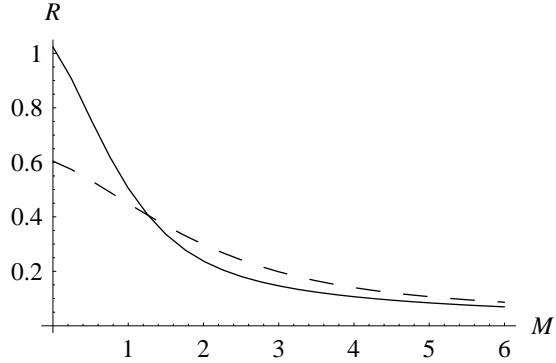}
\end{center}
\caption{Ratio $R=M^2\frac{d\sigma_{gg}}{dM^2\, d^2b}\left( \frac{d\sigma_{gg}}{d^2b}\right)^{-1}$ as a function of invariant mass $M$. Values of parameters for  solid line:  $Q_s=1$~GeV, dashed line: $Q_s=2$~GeV. In both cases $e\mu=1$~GeV.}
\label{Mspec:fig}
\end{figure}

Finally, the total inclusive cross section is determined from \eq{xsect} by first integrating over $\bm \kappa$ which yields the delta function $\delta(\bm \xi-\bm \eta)$. Then using \eq{appc1} and \eq{appc2} we derive
\beq\label{totX}
\frac{d\sigma_{gg}}{d^2b}=\frac{2\as^4 N_cC_F}{3\pi^2}\int_0^\infty \frac{dx}{x}\,2\,\Xi(\b x,\b x)= \frac{16\,\as^4N_cC_F}{3\pi^2}\ln(Q_s/\mu)\,.
\eeq
Numerical calculation of the ratio of \eq{M-spec} to \eq{totX} is exhibited in \fig{Mspec:fig}.

\section{Anomaly matching}\label{sec:3}

Now as we derived the cross section for the double-inclusive gluon production   in the scalar color singlet channel $J^{PC}=0^{++}$, we can use the anomaly matching procedure to derive the double inclusive pion production at low invariant masses. This approach has been discussed in details in \cite{Kharzeev:1999vh}.  Here we give a brief review.

In the product of the ``wave function" and its complex conjugate, there appears a loop formed by the produced gluons. The contribution of this loop is proportional to the correlator\footnote{All coordinate and momenta notations in this section are independent from the notations in other sections unless otherwise specified.}
\beq\label{correl}
\langle 0|T\{ \theta_\mu^\mu(x)\theta_\nu^\nu(0)\}|0\rangle \,,
\eeq
where $\theta^{\mu\nu}$ is the energy-momentum tensor. In the chiral limit its trace acquires a finite value
\beq\label{traceQCD}
\theta_\mu^\mu=-\frac{b\,g^2}{32\pi^2}\,F^{a\nu\rho}F^a_{\nu\rho}\,
\eeq
due to the scale anomaly of QCD. The correlator \eq{correl} can written in the spectral representation
\beq\label{pik2}
\Pi(k^2)=i\int d^4x\, e^{ik\cdot x}\, \langle 0|T\{ \theta_\mu^\mu(x)\theta_\nu^\nu(0)\}|0\rangle =\int d\sigma^2\,\frac{\rho_\theta(\sigma^2)}{\sigma^2-k^2-i0}\,,
\eeq
where the spectral density
\beq\label{rho0}
\rho_\theta(k^2)=\sum_n\int \frac{d^3p_n}{2\,\e_n}(2\pi)^3\,\delta^{(4)}(p_n-k)\,|\langle n|\theta_\mu^\mu|0\rangle|^2\,.
\eeq
$k^2=M^2$ is the invariant mass of the produced system. In the lowest order of perturbation theory the spectral density is given by
\beq\label{rho.pert}
\rho_\theta^\mathrm{pt}(M^2)=\left(\frac{bg^2}{32\pi^2}\right)^2\,\frac{2N_cC_F}{4\pi^2}\,M^4\,.
\eeq

Although the scale invariance of QCD (in the chiral limit) is broken down by quantum fluctuations, there remains a residual symmetry which manifests itself in an infinite tower of  equations, known as the \emph{low energy theorems} \cite{Novikov:1981xj}, relating various Green's functions involving  operator $\theta_\mu^\mu(x)$.  The first in this tower of equations relates the Green's function of the first and second order as follows
\beq\label{pi0}
\Pi(0)=-4\langle 0|\theta_\mu^\mu|0\rangle=-16\,\epsilon_\mathrm{vac}\,.
\eeq
Spectral density \eq{rho0} represents sum over all physical states, perturbative (high $M^2$) and non-perturbative (low $M^2$). We can see using \eq{rho.pert} in \eq{pik2} that the perturbative contribution to  $\Pi(0)$ is divergent. Therefore,   it must be subtracted to satisfy the theorem \eq{pi0}:
\beq\label{subtr}
\int\frac{d\sigma^2}{\sigma^2}[\rho_\theta^\mathrm{phys}(\sigma^2)-\rho_\theta^\mathrm{pt}(\sigma^2)]=-16\,\epsilon_\mathrm{vac}\,.
\eeq
Thus, the vacuum expectation value of $\theta_\mu^\mu$ measures the energy density of non-perturbative fluctuations of vacuum.

If the invariant mass of the produced gluons is small, we can no longer describe the produced particles in terms of the color degrees of freedom. Rather it is appropriate to express $\theta_\mu^\mu$ directly in terms of hadrons. This can be done using the effective chiral Lagrangian \cite{Voloshin:1980zf}
\beq
\lagr=\frac{f^2_\pi}{4}\,\Tr\,\partial_\mu U\,\partial^\mu U^\dagger+\frac{1}{4}m_\pi^2\,f_\pi^2\,\Tr(U+U^\dagger)\,,
\eeq
where $U=e^{2i\pi/f_\pi}$, $\pi\equiv \pi^at^a$ and $t^a$ are the $SU(2)$ generators. The trace of the energy-momentum tensor for this Lagrangian is
\beq\label{trace-pis}
\theta_\mu^\mu=-\partial_\mu\pi^a\,\partial^\mu\pi^a+2\,m_\pi^2\pi^a\pi^a +
\ldots\,.
\eeq
In the chiral limit we have
\beq\label{corr34}
\langle \pi^+\pi^-|\theta_\mu^\mu|0\rangle=M^2\,,
\eeq
leading to the following non-perturbative contribution to the spectral density
\beq\label{rhopipi}
\rho^{\pi\pi}_\theta(M^2)=\frac{3}{32\pi^2}\,M^4\,,
\eeq
where $M^2$ is assumed to be less than a certain cutoff  $M_0$ at which the perturbation theory becomes applicable. This cutoff is related to the vacuum energy density $\epsilon_\mathrm{vac}$ and is estimated to be rather large
$M_0\simeq 2-2.5$~GeV \cite{Novikov:1981xj,Fujii:1999xn}.

\begin{figure}[ht]
    \begin{center}
      \includegraphics[width=6cm]{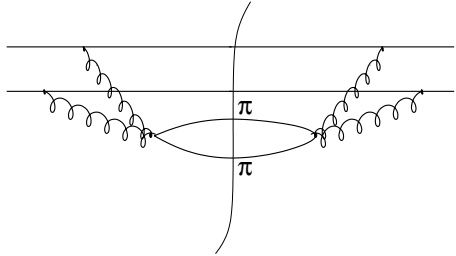}
\end{center}
\caption{Production of a pion pair in $J^{PC}=0^{++}$ channel with invariant mass $M\le M_0$.}
\label{2pions}
\end{figure}

The main idea of anomaly matching is that while at large $M^2$ the physical spectral density coincides with the perturbative formula \eq{rho.pert}, at low $M^2$ it is specified by \eq{rhopipi}. According to \eq{subtr}  gluons do not contribute to the spectral density at low $M^2$.
The two formulas must coincide at the scale $M_0$. Therefore, in order to  calculate  production of two pions at $M\le M_0$ we need to replace the perturbative contribution to the spectral density \eq{rho.pert} --- calculated (indirectly) in the previous section --- by the two-pion contribution \eq{rhopipi}.
This substitution is equivalent to calculating the  diagram in \fig{2pions}.
Therefore, the cross section for the double-inclusive pion production becomes (using notation of the previous section)
\bea\label{main}
\frac{d\sigma_{\pi\pi}}{dM^2\,d^2k\,dy\,d^2b}&=&\frac{96\,\as^2}{\pi^5\,b^2}\,\frac{1}{M^4}\,
\int \frac{d^2x}{x^2}\int\frac{d^2y}{y^2}\, e^{i\b k\cdot (\b x-\b y)}
\,\nonumber\\
&\times&
\big[1-ixM\, K_1(i x M)\big]\,\big[1+iyM\, K_1(-i y M)\big]
\, 2\,\Xi(\b x, \b y)\,,
\eea
where we employed \eq{rho.pert} and \eq{rhopipi} in \eq{XSECT3}.
It is remarkable that this cross section is of the order $\as^2$, two remaining factors of $\as$ are inherent to  the gluon distribution functions of each projectile nucleon.

\section{Including quantum evolution}\label{sec:evol}

Here we are going to include the small-$x$ nonlinear quantum evolution of \cite{BK} into the cross section from Eq.~\eq{XSECT3}. Since the evolution equations in \cite{BK} are written for the forward amplitude of a quark dipole on a nucleus, we have to first generalize
Eq.~\eq{XSECT3} to the case of gluon pair production in two dipoles--nucleus scattering. Of course, such a model of a nucleon is a rough approximation. However, in the gluon saturation regime details of  nucleon structure play a little role. Therefore, our results below may still serve as a good approximation to a more accurate treatment of the nucleon \cite{Kovchegov:2001sc}. The generalization of \eq{XSECT3} to dipole-nucleus scattering is easily done by
including emissions of the $s$-channel gluon in \fig{wave_funct} by the quark and anti-quark in the incoming dipoles. Denote the transverse coordinates of the quark and anti-quark in the incoming dipoles by $^a\b z_0$ and $^a\b z_1$, where the superscript $a=1,2$ labels different nucleons. Then, instead of  \eq{XSECT3} we have
\bea\label{xdip}
&&\frac{d\sigma_{gg}}{dM^2\,d^2k\,dy\,d^2b}(^1\b z_{01},^2\b z_{01})=\frac{8\as^4}{\pi^7}\,\frac{N_cC_F}{M^4}\,\int \frac{d^2x}{x^2}\int\frac{d^2y}{y^2}\, e^{i\b k\cdot (\b x-\b y)}\, \sum_{a=1}^2\sum_{i,j=0}^1\,2\,\Xi(\b x,\b y; {^a\b z_i}, {^a\b z_j})\nonumber\\
&&\quad \times
 \bigg[1-i\big|\b x- {^a\b z_i}\big|M\, K_1\big(i \big|\b x- {^a\b z_i}\big| M\big)\bigg]\,\bigg[1+i\big| \b y-{^a\b z_j} \big|M\, K_1\big(-i \big| \b y- {^a\b z_j} \big| M\big)\bigg]
\,.
\eea
where $^a\b z_{01}={^a\b z_0}-{^a\b z_1}$ and
\beq\label{scat-dip}
\Xi(\b x,\b y;{^a\b z_i},{^a\b z}_j)=\left(e^{-\frac{1}{4}({^a\b z_i}-{^a\b z_j})^2Q_s^2}+e^{-\frac{1}{4}(\b x-\b y)^2 Q_s^2}
 - e^{-\frac{1}{4}(\b x-{^a\b z_i})^2 Q_s^2}-e^{-\frac{1}{4}(\b y-{^a\b z_j})^2 Q_s^2}\right)^2\,.
 \eeq

The inclusion of quantum corrections in the leading logarithmic
 approximation (resumming powers of $\as \, y$) in the large-$N_c$
limit is done along the lines of \cite{Kovchegov:2001sc}
using Mueller's dipole model formalism \cite{dipole}. Since we assume that the produced gluons are at the same rapidity, the prescription
for inclusion of quantum evolution is identical to the single gluon
production case. We first define the quantity $n_1 ({^a\b z_0}, {^a\b
z_1}; {^a\b w_0}, {^a\b w_1}; Y-y)$, which has the meaning of the
number of dipoles with transverse coordinates ${^a\b w}_{0}, {^a\b
w_{1}}$ at rapidity $y$ generated by the evolution from the original
dipole ${^a\b z}_{0}, {^a\b z_1}$ having rapidity $Y$. It obeys the dipole
equivalent of the BFKL evolution equation \cite{dipole,BFKL}
\bea\label{eqn}
\frac{\partial n_1 ({^a\b z_0}, {^a\b z_1}; {^a\b w_0}, {^a\b w_{1}}; y)}{\partial y} & = &
\frac{\as \, N_c}{2 \, \pi^2} \,
\int d {^a\b z_2} \, \frac{z_{01}^2}{z_{20}^2 \, z_{21}^2} \,
\bigg[ n_1 ({^a\b z_0}, {^a\b z_2}; {^a\b w_0}, {^a\b w_1}; y)
 \nonumber\\
&&
+n_1 ( {^a\b z_2}, {^a\b z_1}; {^a\b w_0}, {^a\b w_{1}}; y)- n_1 ({^a\b z_0}, {^a\b z_1}; {^a\b w_0}, {^a\b w_{1}}; y) \bigg]
\eea
with the initial condition
\beq
n_1 ({^a\b z_0}, {^a\b z_1}; {^a\b w_0}, {^a\b w_{1}}; y=0) \, = \,
\delta ( {^a\b z_0} - {^a\b w_{0}} ) \, \delta ( {^a\b z_1} - {^a\b w_{1}} ).
\eeq
If the target nucleus has rapidity $0$, the incoming dipole has
rapidity $Y$, and the produced gluons have rapidity $y$, the inclusion
of small-$x$ evolution in the rapidity interval $Y-y$ is accomplished
by replacing the cross section from \eq{xdip} by \cite{Kovchegov:2001sc,Jalilian-Marian:2005jf}
$$
\frac{d \, \sigma_{gg}}{dM^2\,d^2k\,dy\,d^2b} ({^1\b z_{01}},{^2\b z_{01}})
\rightarrow
$$
\beq\label{sub1}
 \int\prod_{a=1}^2 d {^a\b w_{0}} \, d {^a\b w_{1}} \, n_1 ({^a\b z}_{0}, {^a \b z_1};
{^a\b w}_{0}, {^a\b w_{1}}; Y-y) \,
\frac{d \, \sigma_{gg}}{dM^2 \, d^2 k \, dy  \, d^2 b} ({^1\b w_{01}},{^2\b w_{01}})\,.
\eeq
Eq.~\eq{sub1} neglects correlations between the original dipoles as explained in Sec.~\ref{sec:intr}.
 The evolution in each of the original dipoles is   linear as was originally shown in \cite{Kovchegov:2001sc}: the Pomeron splittings cancel in the rapidity interval between $y$ and $Y$ in compliance with the AGK cutting rules \cite{Abramovsky:1973fm}.

Inclusion of evolution in the interval between $0$ and $y$ is
accomplished by replacing the Mueller-Glauber rescattering exponents
according to the following rule \cite{Kovchegov:2001sc}
\beq\label{sub2}
e^{- \frac{1}{4} \, (\b x_0 - \b x_1)^2 \, Q_s^2} \,  \, \rightarrow \,
1 - N (\b x_0, \b x_1, Y)\,,
\eeq
where $N (\b x_0, \b x_1, Y)$ is the forward amplitude for a quark
dipole $\b x_0, \b x_1$ scattering on a target with rapidity
interval $Y$ between the dipole and the target. It obeys the following
evolution equation \cite{BK}
$$
\frac{\partial N ({\b x}_{0}, {\b x}_1, Y)}{\partial Y} \, = \,
\frac{\as \, N_c}{2 \, \pi^2} \,
\int d^2 x_2 \, \frac{x_{01}^2}{x_{20}^2 \, x_{21}^2} \,
\left[ N ({\b x}_{0}, {\b x}_2, Y) +
N ({\b x}_{2}, {\b x}_{1}, Y) - N ({\un
x}_{0}, {\b x}_1, Y) \right.
$$
\beq\label{eqN}
- \left. N ({\b x}_{0}, {\b x}_{2}, Y) \, N ({\b x}_{2}, {\b x}_{1}, Y)
\right]
\eeq
with the initial condition
\beq
 N ({\b x}_{0}, {\b x}_1, Y=0) \, = \, 1 - e^{- \frac{1}{4} \, (\b
 x_0 - \b x_1)^2 \, Q_s^2 }\,.
\eeq

Performing the substitution from \eq{sub2} in \eq{scat-dip} yields
\beq\label{xi_ev}
\Xi(\b x,\b y;{^a\b z_i},{^a\b z};Y)=
\left(
N(\b x,{^a\b z_i},Y)+N(\b y,{^a\b z_j},Y)-N({^a\b z_i},{^a\b z_j},Y)-N(\b x,\b y,Y)
 \right)^2\,.
\eeq

With the definition of Eqs. (\ref{xi_ev}) we write the following
answer for the double inclusive gluon production cross section in the scalar $J^{PC}=0^{++}$ channel
including small-$x$ evolution effects
$$
\frac{d  \sigma_{gg}}{dM^2\,d^2k\,dy\,d^2b} (^1\b z_{01},^2\b z_{01})
 =  \frac{8\as^4}{\pi^7}\,\frac{N_cC_F}{M^4}\,
\int \prod_{a=1}^2 d{^a \b w_{0}} \, d {^a\b w_{1}} \, n_1 ({^a\b z}_{0}, {^a\b z_1};
{^a\b w}_{0}, {^a\b w_{1}}; Y-y)
$$
$$
\times\,\int \frac{d^2x}{x^2}\int\frac{d^2y}{y^2}
 \, e^{i\b k\cdot (\b x-\b y)}\, \sum_{i,j=0}^1\,2\,\Xi(\b x,{^a\b w_i}; \b y,{^a\b w_j};y)\,
 $$
 \beq\label{dcl_ev}
 \times\,\big[1-i|\b x- {^a\b w_i}|M\, K_1(i |\b x- {^a\b w_i}| M)\big]\,\big[1+i| \b y-{^a\b w_j} |M\, K_1(-i | \b y-{^a\b w_j} | M)\big]\,.
\eeq
Similarly, the double inclusive pion production cross section is given using \eq{main} by
$$
\frac{d  \sigma_{\pi\pi}}{dM^2\,d^2k\,dy\,d^2b} (^1\b z_{01},^2\b z_{01})
 =  \frac{96\,\as^2}{\pi^5\,b^2}\,\frac{1}{M^4}\,
\int \prod_{a=1}^2 d{^a \b w_{0}} \, d {^a\b w_{1}} \, n_1 ({^a\b z}_{0}, {^a\b z_1};
{^a\b w}_{0}, {^a\b w_{1}}; Y-y)
$$
$$
\times\,\int \frac{d^2x}{x^2}\int\frac{d^2y}{y^2}
 \, e^{i\b k\cdot (\b x-\b y)}\, \sum_{i,j=0}^1\,2\,\Xi(\b x,{^a\b w_i}; \b y,{^a\b w_j};y)\, $$
 \beq\label{main_ev}
 \times\,\big[1-i|\b x- {^a\b w_i}|M\, K_1(i |\b x- {^a\b w_i}| M)\big]\,\big[1+i| \b y-{^a\b w_j} |M\, K_1(-i | \b y-{^a\b w_j} | M)\big]\,.
\eeq
This is the central result of our paper. We are going to use it for the phenomenological analysis of the RHIC data in the forthcoming publication \cite{FP}.



\vskip0.3cm
{\large\bf Acknowledgments.}
We acknowledge a very stimulating and informative discussions with Dima Kharzeev, Yuri Kovchegov and Jianwei Qiu. K.T.  would like to thank
RIKEN, BNL and the U.S. \!\! Department of Energy (Contract
No.\!\! DE-AC02-98CH10886) for providing the facilities essential for the
completion of this work.


\appendix
\section{Derivation of the inclusive cross section in the $z,z'=1/2$ approximation }

Using expressions   \eq{wfA2}, \eq{F}  the  function $F(\bm\xi,\b x)$  becomes
\beq\label{F12}
F(\bm \xi,\b x)=
\int\frac{d^2k}{(2\pi)^2}\int\frac{d^2\kappa}{(2\pi)^2}e^{-i\,\bm \kappa\cdot \bm \xi-i\,\b k\cdot \b x}
\int \frac{d^2 q}{16\pi^3} 4\,\frac{(\frac{1}{2}\,\b k-\b q)\cdot (\frac{1}{2}\,\b k+\b q)}{(\frac{1}{2}\,\b k-\b q)^2\,(\frac{1}{2}\,\b k+\b q)^2}\,
 \frac{1}{\b q^2 \,  - \bm\kappa^2-i0}\,.
\eeq
Integration over $\b k$ can be done using the following formula (see Appendix B for the derivation)
\beq\label{FT}
\int\frac{d^2k}{(2\pi)^2}\frac{(\frac{1}{2}\,\b k-\b q)\cdot (\frac{1}{2}\,\b k+\b q)}{(\frac{1}{2}\,\b k-\b q)^2\,(\frac{1}{2}\,\b k+\b q)^2}\,e^{-i\,\b k\cdot \b x}\,=\,-
\frac{1}{\pi}\frac{\b q\cdot \b x}{\b q^2\,\b x^2}\, \sin(2\,\b q\cdot \b x)\,.
\eeq
The subsequent integration over $\bm \kappa$ is  a Fourier transformation of a two-dimensional Feynman propagator and can be expressed in terms of the
Hankel function
\beq\label{Fourier.prop}
\int \frac{d^2 \kappa}{(2\pi)^2}\,e^{-i\bm \kappa\cdot \bm \xi}\,\frac{1}{\b q^2-\bm \kappa^2 - i0}=\frac{i}{4}\,H_0^{(2)}(q\,\xi)\,,
\eeq
where $q=|\b q|$, $\xi=|\bm \xi|$. With the aid of \eq{FT} and \eq{Fourier.prop} we obtain
\beq
F(\bm \xi, \b x)=\frac{i}{8\pi^3x}\int_0^\infty dq\, H_0^{(2)}(q\xi)\,J_1(2qx)\,.
\eeq
This integral can be taken  by analytically continuing the Hankel function to imaginary values of  $q$  with the help of  Eq.~9.6.4 of Ref.~\cite{AS}
\beq
K_0(ir)=-\frac{i\pi}{2}\,H_0^{(2)}(r) \,,
\eeq
and using Eq.~6.576.3 of Ref.~\cite{GR}. The result is
\beq\label{r1}
F(\bm \xi, \b x)=-\frac{1}{(2\pi)^4\, x^2}\,\ln\left( 1-\frac{4x^2}{\xi^2}\right) \,.
\eeq

To calculate the cross section \eq{xsect} we use the following integral
\bea\label{i1}
&&\int d^2\xi\, e^{-i\frac{1}{2}\,\bm \xi\cdot(\b l_2-\b l_1)}\,\ln\left( 1-\frac{4x^2}{\xi^2}\right) =2\pi\int_0^{\infty} d\xi\,\xi \, J_0\left(\frac{1}{2}\xi\,|\b l_2-\b l_1|\right)\,
\ln\left(\frac{4x^2}{\xi^2}-1\right) \nonumber\\
&=&\frac{16\,\pi}{(\b l_2-\b l_1)^2}\big[1-ix|\b l_2-\b l_1|\, K_1(i x |\b l_2-\b l_1|)\big]\,.
\eea
Inserting \eq{r1} and \eq{i1} into \eq{xsect}  and using $K_1^*(iz)=K_1(-iz)$ we derive \eq{xsect2}.

Let us also note for future reference the following integral
\bea\label{appc2}
&&\int_0^\infty d\xi\, \xi \left| \ln\left(1-\frac{4x^2}{\xi^2}\right)\right|^2
 =\int_0^{2x}d\xi\,\xi\left[\ln^2\left(\frac{4x^2}{\xi^2}-1\right)+\pi^2\right]+
\int_{2x}^\infty d\xi\, \xi\ln^2\left(1-\frac{4x^2}{\xi^2}\right)\nonumber\\
&&=\frac{2\pi^2x^2}{3}+2\pi^2x^2-\frac{4\pi^2x^2}{3}=\frac{4\pi^2x^2}{3}\,.
\eea
\section{Derivation of \eq{FT}}

Consider an auxiliary function $G(\b x, \b q)$ defined as following
\beq\label{Gf}
G(\b x, \b q)=\int\frac{d^2k}{(2\pi)^2}\,e^{-i\b k\cdot\b x}\,\frac{1}{(\frac{1}{2}\,\b k-\b q)^2\,(\frac{1}{2}\,\b k+\b q)^2}\,.
\eeq
One can readily verify that
\beq\label{tr1}
\int\frac{d^2k}{(2\pi)^2}\frac{(\frac{1}{2}\,\b k-\b q)\cdot (\frac{1}{2}\,\b k+\b q)}{(\frac{1}{2}\,\b k-\b q)^2\,(\frac{1}{2}\,\b k+\b q)^2}\,e^{-i\,\b k\cdot \b x}\,=\,
-\left(\frac{1}{2}\nabla_x+i\b q\right)\left(\frac{1}{2}\nabla_x-i\b q\right)\,G(\b x, \b q)\,.
\eeq
Thus, the problem is reduced to  evaluation of $G(\b x, \b q)$. Using the Feynman's trick we write
$$
G(\b x, \b q)=\int_0^1d\alpha\int\frac{d^2k}{(2\pi)^2}\,e^{-i\b k\cdot\b x}\,
\frac{1}{\{\alpha(\frac{1}{2}\b k-\b q)^2+(1-\alpha)(\frac{1}{2}\b k+\b q)^2\}^2 }\,.
$$
Introducing a new vector $\b k'=\frac{1}{2}\b k+(1-2\alpha)\,\b q$ we come by
\bea\label{hj1}
G(\b x, \b q)&=& 4\int_0^1d\alpha\, e^{2(1-2\alpha)\,i\b q\cdot \b x}\int\frac{d^2k'}{(2\pi)^2}e^{-2i\b k'\cdot \b x}\frac{1}{\{{\b k'}^2+4\,\alpha(1-\alpha) \b q^2\} ^2}\nonumber\\
&=& \frac{1}{\pi}\int_0^1d\alpha\, e^{2(1-2\alpha)\,i\b q\cdot \b x}\,
\frac{x}{q\sqrt{\alpha(1-\alpha)}}\,K_1(4xq\sqrt{\alpha(1-\alpha)})\,.
\eea

Now, the integral \eq{Gf} is dominated by two IR logarithmic singularities at $\b k=\pm 2\b q$ which tie in with emission of soft gluons $\b k_1=0$ and $\b k_2=0$ by the projectile. In \eq{hj1} they correspond to the values $\alpha=0,1$ of the Feynman parameter $\alpha$. Keeping only the logarithmically divergent
terms we derive
\beq
G(\b x, \b q)=\frac{x}{\pi\,q}\left( e^{2i\b q\cdot\b x}+e^{-2i\b q\cdot\b x}\right)
\int_0^\infty\frac{d\alpha}{\sqrt{\alpha}}\,K_1(4xq\sqrt{\alpha})\,.
\eeq
Changing the integration variable $\alpha=\beta^2$ yields
\beq\label{pop}
G(\b x, \b q)= \frac{4x}{\pi q}\, \cos(2\b q\cdot\b x)\,\lim_{\delta\to 0}\int_0^\infty d\beta\, \beta^\delta\, K_1(4xq\beta)
= \frac{1}{\pi q^2}\, \cos(2\,\b q\cdot\b x)\,\ln(1/qx)\,.
\eeq
Finally, using \eq{pop} in \eq{tr1} we arrive at \eq{FT}.

An important remark is in order here. Eq.~\eq{FT} is not  valid in two limits: (i) $\b x\to 0$, $\b q$ fixed and (ii) $\b q\to 0$, $\b x$ fixed, though it holds in the limit $q x\to 0$.   The reason is that the integral on the  left-hand-side of \eq{FT}
is symmetric with respect to transformation $\b x\leftrightarrow \b q$. This can be verified using twice the formula
\beq
\frac{\b k}{\b k^2}=\int d^2z\,e^{i\b k\cdot\b z}\,\frac{1}{2\pi i}\,\frac{\b z}{\b z^2}
 \eeq
to transform the integrand into the coordinate space. This symmetry property is violated in the above mentioned limiting cases (i) and (ii). The values of the integral in these cases are
$$
\int\frac{d^2k}{(\frac{1}{2}\b k)^2}\,e^{-i\b k\cdot \b x}=8\pi \ln(1/x\mu)
$$
and \cite{Kovchegov:1998bi}
$$
\int d^2k\,\frac{(\frac{1}{2}\,\b k-\b q)\cdot (\frac{1}{2}\,\b k+\b q)}{(\frac{1}{2}\,\b k-\b q)^2\,(\frac{1}{2}\,\b k+\b q)^2}=-8\pi\,\ln(q/\mu)\,,
$$
where 
$\mu$ is an IR cutoff.

\section{Convergence of the integral in \eq{xsect3}}

Consider the following integral
$$
I(M,k) =
\frac{1}{(2\pi)^2}\int \frac{d^2x}{x^2}\int\frac{d^2y}{y^2}\, e^{i\b k\cdot (\b x-\b y)}
\big[1-ixM\, K_1(i x M)\big]\,\big[1+iyM\, K_1(-i y M)\big]
\,\Xi(\b x, \b y)\,.
$$
The dipole scattering amplitude $\Xi(\b x,\b y)\le 1$ by unitarity. Therefore,
$$
I(M,k)\le \left|\int \frac{d^2x}{(2\pi)\,x^2}\,e^{i\b k\cdot\b x}\,\big[1-ixM\, K_1(i x M)\big]\right|^2 = \left|\int_0^\infty \frac{dx}{x}\, J_0(k x)\,\big[1-ixM\, K_1(i x M)\big]\right|^2\,,
$$
where $x=|\b x|$ and $k=|\b k|$ as usual.
Since (see \eq{i1})
\beq\label{appc1}
\int_0^\infty \frac{dx}{x}J_0(k x)\,\big[1-ixM\, K_1(i x M)\big]=\ln\left(\frac{M^2}{k^2}-1\right)\,,
\eeq
we derive
\beq
I(M,k)\le \left|\ln\left(\frac{M^2}{k^2}-1\right)\right|^2\,.
\eeq
This expression is finite apart from the logarithmic divergence at $k=M$ corresponding to $\b l_1\cdot \b l_2=0$. Integration over $M$ in \eq{XSECT3}
is thus bounded by (see \eq{appc2})
\beq
\int_0^\infty\frac{dM}{M^3}\,I(M,k)\,\le \int_0^k\frac{dM}{M^3}\,
\ln^2\left(1-\frac{M^2}{k^2}\right)+
\int_k^\infty\frac{dM}{M^3}\,\left[
\ln^2\left(\frac{M^2}{k^2}-1\right)+\pi^2\right]=\frac{5\pi^2}{6\,k^2}\,.
\eeq


\end{document}